# Spreadsheet Debugging

Yirsaw Ayalew
Department of Computer Science
Addis Ababa University
P.O.Box 1176
Addis Ababa, Ethiopia
yirsawa@math.aau.edu.et

Roland Mittermeir
Institut of Informatics-Systems
University of Klagenfurt
Universitätsstr. 65-67,
A-9020, Klagenfurt, Austria
roland@isys.uni-klu.ac.at

**ABSTRACT**

Spreadsheet programs, artifacts developed by non-programmers, are used for a variety of important tasks and decisions. Yet a significant proportion of them have severe quality problems. To address this issue, our previous work presented an interval-based testing methodology for spreadsheets. Interval-based testing rests on the observation that spreadsheets are mainly used for numerical computations. It also incorporates ideas from symbolic testing and interval analysis.

This paper addresses the issue of efficiently debugging spreadsheets. Based on the interval-based testing methodology, this paper presents a technique for tracing faults in spreadsheet programs. The fault tracing technique proposed uses the dataflow information and cell marks to identify the most influential faulty cell(s) for a given formula cell containing a propagated fault.

## 1. INTRODUCTION

Spreadsheet systems are widely used and highly popular end-user systems. They are used for a variety of important tasks such as mathematical modeling, scientific computation, tabular and graphical data presentation, data analysis and decision-making. Many business applications are based on the results of spreadsheet computations and consequently important decisions are made based on spreadsheet results. As we cannot assume professionalism of spreadsheet developers in writing and testing their spreadsheet programs, we cannot assume professionalism in debugging.

Therefore, this paper presents an approach to help users in identifying cells containing actually faulty entries. The unifying property of spreadsheet applications is that they involve numeric computations. Numeric computations constitute the primary turf of spreadsheets, in spite of the fact that the spreadsheet model finds derived applications in other areas too. These are as diverse as information visualization [7], concurrent computation [22, 23], or user interface specifications [10], to name just a few. There is also a trend towards using the spreadsheet model as a general model for end-user programming [14].

Despite their popularity due to their ease of use and suitability for numerical computations, a significant proportion of spreadsheet programs have severe quality problems. In recent years, there has been an increased awareness of the potential impact of faulty spreadsheets on business and decision-making. A number of experimental studies and field audits [5, 9, 15, 16, 17, 18, 19, 20] have shown the serious impact spreadsheet errors have on business and on decisions made on the basis of results computed by spreadsheets.

In contrast to the professional use of the results computed by spreadsheets, the development of these results is less professional. The developers of spreadsheets are mainly end users who are not expected to follow a formal process of software development although the sheets they produce are of the nature of regular programs and quite often reach the complexity of typical data processing application software[1]. Nevertheless, spreadsheet developers rather uncritically rely on the initial correctness of their programs.

To address this problem, our previous work [3, 4, 13, 24] presented an interval-based testing methodology for spreadsheets. The interval-based testing methodology was proposed based on the premise that spreadsheet developers are not software professionals. The approach takes into account inherent characteristics of spreadsheets as well as the conceptual models of spreadsheet programmers. It also incorporates ideas from symbolic testing and interval analysis. Interval-based testing focuses on the functionality of spreadsheet formulas instead of the internal structure of a spreadsheet program (i.e., it is not based on a traditional code coverage criterion).

With interval testing, cells containing suspicious values are identified. But as with conventional programming, the spot where a wrong value figures is not necessarily the location of the fault in the program. This might be some step in the algorithm that has been executed on the way to the respective output statement. Since with spreadsheets, basically all intermediate results of computations are visible on the user interface (exceptions are only computations within a cell and computations in hidden cells), the issue of directing spreadsheet users from the cell containing a wrong result to the cell containing the wrong formula becomes specifically important.

Based on the interval-based testing methodology, this paper presents a technique for tracing faults in a spreadsheet program. Cells whose value is outside a computed interval (see section 2) are referred to as cells containing a symptom of fault. The issue is, given a formula cell with a symptom of fault which depends directly and/or indirectly on other cells which have symptoms of faults, how to identify the most influential faulty cell. If we can identify the most influential faulty cell, then correcting that cell may correct many of the cells showing incorrect values, which are dependent on it thereby simplifying the debugging process. The fault tracing approach uses information from the interval-based testing system about the verification status of the cells, dataflow information, and priority values of cells.

---

[1] For this reason, we refer to the system of interlinked data and formula cells of a spreadsheet also as "spreadsheet program"

This paper is organized as follows: Section 2 briefly describes the interval-based testing methodology. Section 3 discusses the general problem of fault tracing in relation to conventional software debugging techniques and spreadsheet debugging. Section 4 discusses the fault tracing strategy supported by the tool developed to support interval-based spreadsheet analysis. While tracing for faults, there is a need for minimizing the search region. This requires the identification of those faulty cells, which have a higher likelihood of containing the most influential faulty cell(s). This is presented in Section 4.1. An example and a fault tracing algorithm are described in this same section. Finally, the main points of the fault tracing approach are summarized in Section 5.

## 2. INTERVAL-BASED TESTING

Generally, the main task in testing a program is to be able to detect the existence of symptoms of faults in the program. By running the program with test cases and comparing the result with the expected outcome described in the specification or generated by a test oracle, the existence of a fault can be detected. However, in spreadsheet programming most spreadsheet programmers do not have the expertise to design and execute effective test cases. In the absence of a specification and automated test oracle, the user plays the role of a test oracle and provides the expected outcome during testing.

The root of spreadsheet programming lies in the definition of formulas. In spreadsheets, users want to make sure that their spreadsheet formulas are correct with respect to the actual data that they need for their applications instead of arbitrary data chosen for testing purposes (this does not include those who develop templates). Hence, the reasonableness of the computed value is used to judge the validity of the formula. Users usually have a gut feeling of the range of reasonable values for each given cell.

Interval-based testing is proposed to check the existence of symptoms of faults in formulas, which are defined for numerical computations. Spreadsheets are mainly used for numerical computations by end users. Hence, we require from the user a vision of the ranges of possible values of formula computations.

Interval-based testing requires the user to specify for a given spreadsheet on a mirror image of the sheet intervals expected for the desired input and formula cells respectively. For numeric input cells, the user specifies the range of reasonable values in the form of intervals, which serve as input domains. For formula cells, the user specifies the expected outcome of the formula again in the form of intervals. The prototype system developed in the context of [3], an add on to spreadsheet packages like MS-Excel, allows also for selective specification of such expected intervals. It mirrors the user's sheet by means of an interval-value sheet containing boundaries within which the spreadsheet user expects the result of a formula to be. Further, an interval-formula sheet where such boundaries are derived based on the formulas of the users actual sheet by means of interval arithmetic applied to the intervals given for input cells is established by the system. The attached intervals will be stored as strings (since the spreadsheet system used with the prototype neither supports interval data types nor allows user defined data types) in a behind-the-scene spreadsheet using the same cell coordinates as the cells in the ordinary spreadsheet.

*Figure 1: Architecture of the prototype supporting the interval-based testing methodology*

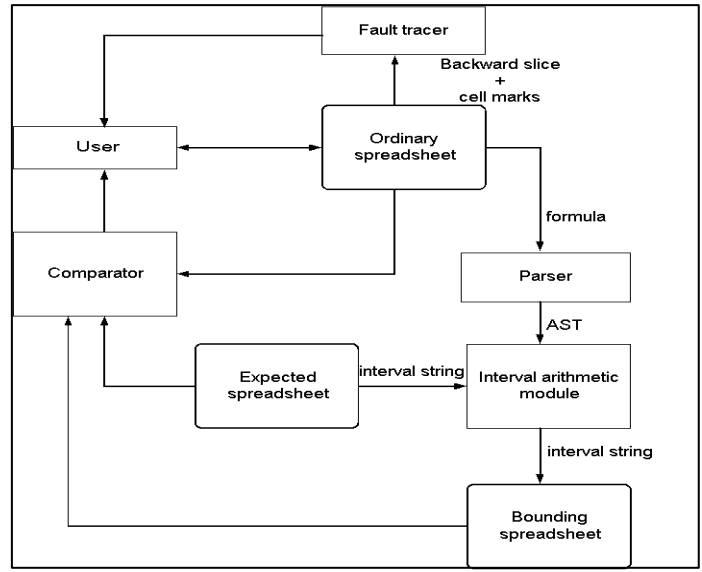

In figure 1, ***Ordinary spreadsheet*** is the usual spreadsheet defined by the user in which computation is based on discrete values. ***Expected spreadsheet*** is a behind-the scene spreadsheet which contains expected intervals for input and formula cells as given by the user. For a formula defined at the ordinary spreadsheet, the formula is evaluated based on the interval values stored in the expected spreadsheet and the resulting interval will be stored as an interval string in the respective cell in the bounding spreadsheet. In some cases, a user may not attach input and expected intervals for some input and formula cells. In such situations, the discrete values of the cells from the ordinary spreadsheet are used during interval computation as intervals of width zero. The ***bounding spreadsheet***, which is a behind-the-scene spreadsheet, contains computed bounding interval values. A bounding interval is an interval computed by the formulas in the users regular spreadsheet with the operators in formula cells replaced by the respective interval operators. It is used to check the reasonableness of the expected interval specified by the user. Once the necessary values are available from the three sources, namely, spreadsheet computation, user expectation, and interval computation, the comparator may determine the existence of symptoms of faults. Whenever there is a discrepancy between spreadsheet computation, user expectation and interval computation, the comparator marks those cells with symptoms of faults and those which seem to be correct in different colors. Finally, among those cells with symptoms of faults which contribute to a faulty cell, the most influential cell is identified using the fault tracer. A discussion of the fault tracer is the main subject of this paper.

## 3. FAULT TRACING BACKGROUND

Once symptoms of faults are detected, the next task is to find the location of the actual faults. A testing system cannot exactly indicate the location of faults; it rather provides a hint or a symptom of a fault. However, a testing system can facilitate the search for the location of faults by providing testing information about the possible paths that lead to the likely fault location. A symptom of fault is a signal indicating the existence of a possible fault. A symptom of fault is generated whenever there is a discrepancy between the expected behavior of a program and its actual behavior. Fault tracing is the process of identifying the location of faults in a program. In the following sections, we discuss the process of debugging conventional software and spreadsheet debugging.

### 3.1 Debugging conventional software

Generally, fault tracing in conventional software debugging involves program slicing techniques to minimize the search for the potential faults [1, 2, 8, 11].

The first important information needed in fault tracing is to compute a static backward slice. It contains all variables that may affect the variable at which a symptom of fault is detected at a given statement. Since a symptom of fault is generated based on a particular test case, those variables, which are directly and/or indirectly involved in the current computation will contain the statement producing the faulty variable provided that the fault is not due to missing statements. This requires the computation of dynamic backward slices. In order to reach the potential faulty variables, further reduction of the dynamic backward slice should be made using the technique of dicing.

Dicing is carried out by removing the sub-slice corresponding to correct variables. However, the use of dicing imposes some preconditions to be satisfied in order for the resulting dice to contain the fault [12]. For example, dicing assumes that only one faulty variable exists in the dynamic backward slice. In addition, it assumes that if a variable is faulty then all variables in the dynamic forward slice of that variable are faulty. This misses situations where faults compensate within the slice. Hence, the general process of fault tracing in conventional software can be described as follows.

*Static slice ® Dynamic slice ® Dice ® Potential location of fault*

### 3.2 Spreadsheet debugging

The problem in spreadsheet debugging is: given a formula cell with a symptom of fault but not having a fault in the formula, how to identify the faulty cell among those cells on which the given cell directly and/or indirectly depends. Since this analysis shows only cells with deviations between discrete and interval computations, we can only cautiously speak of symptoms of faults. The deviating cells might well result from propagations. In some cases, a symptom of fault can be generated even though there is no fault in the formula. This happens when the user expectation is specified incorrectly and due to propagation of faulty values.

As the conceptual view of a spreadsheet program is based on data dependency relations, spreadsheets can be considered as dataflow driven. On the other hand the conceptual view of a procedural program is control flow-driven. In spite of this difference, a similar procedure to conventional software debugging can be used for fault tracing in spreadsheets. Therefore, the notion of a slice is not directly applicable. We have rather a set of cells linked just by dataflow connection. Nevertheless, for the sake of comparison, we refer to this set as slice. However, in our approach we do not impose the requirement that the dynamic backward slice contains only one faulty cell. There can be several cells in the dynamic backward slice of a faulty cell which are marked as faulty. Cell marking is performed based on the result of the comparison made by the comparator as described in section 2. The fault tracing procedure uses the backward slice and the cell marks recorded by the interval-based testing system.

The fault localization technique, proposed by DeMillo et al. [8], was based on the analysis of the steps used by programmers experienced in debugging. Following a similar procedure, the spreadsheet fault tracing process contains the following steps:

1. Determine the cells (directly and/or indirectly) involved in the computation of an incorrect formula (i.e., look backward)
2. Select suspicious cells
3. Form hypotheses about suspicious cells

Step 1 requires the computation of "backward slices" with respect to a faulty cell for which we want to locate the source of the fault. Step 2 requires the identification of those cells, which have a likelihood of propagating faults through the data flow. These cells are marked by the comparator during the verification process. The comparator determines the existence of a symptom of a fault for a formula cell by comparing the usual spreadsheet computation, user expectation, and interval computation (see figure 1). Step 3 requires reasoning about the most influential cell. In other words, this step involves the computation of the priority values for the suspicious cells and the identification of the one, which has the highest likelihood of contributing to the faulty cells in the dynamic backward slice.

If the fault is local, i.e., either the formula or the expected interval of the given cell is specified incorrectly, then the fault can be fixed by examining the faulty cell itself. If the fault is not local, we are looking for the most influential cell. The assumption in this approach is that if the most influential faulty cell is found, then correcting this cell may correct many of the dependent faulty cells in the data dependency graph, thereby reducing the remaining task of the spreadsheet debugging process. To address this problem, we propose an approach using priority setting based on the number of incorrect precedents and dependents. To do so, we rely on the dynamic backward slice of a given faulty cell as we are dealing with propagated faults.

A similar approach was proposed by Reichwein et al. [21] for debugging Form/3 programs. In this approach, a user marks cells as correct and incorrect and based on the all-uses dataflow test adequacy criterion, the degree of testedness of formulas is compu-ted. Cells are given different colors based on their degree of testedness. Fault tracing is performed based on the degree of testedness of cells and by computing the fault likelihood of cells. Fault likelihood is computed based on the number of correct and incorrect dependents of a cell. During the fault localization process, for the cell under consideration, the cells in the dynamic backward slice will be highlighted in different colors based on their degree of testedness. The further process of fault localization is carried out by performing testing using additional test cases. This approach requires the user to provide different test cases to localize the faulty cell. The work proposed by Chen and Chan [6] presented a model for spreadsheet debugging. This work described the cognitive aspects of spreadsheet debugging and provided the essential episodes, which could be applied to the debugging process in spreadsheets.

Our approach does not require the user to provide different test cases since we cannot assume a sufficient testing discipline. As users are working on actual data that they need for their applications, it is likely that they need to know the correctness of their spreadsheets based on the actual data instead of arbitrary data chosen just for testing purposes. The fault tracing approach presented here requires the computation of priority values based on the verification status of precedent cells. If this is not sufficiently discriminating the priority values are also based on dependent cells.

## 4. FAULT TRACING STRATEGY

This section describes how the fault tracing approach works and presents an example to illustrate the approach. Finally, a fault tracing algorithm is provided.

The fault tracing strategy proposed uses information from different sources to locate the most influential faulty cell. The first information that we need is the dataflow information. This information is already available since it is used by the spreadsheet system during the evaluation of formulas. For example, in Microsoft Excel, this information is used by its built-in auditing tool to show the backward and forward slices of cells of interest one level at a time. While traversing the backward slice, we need a mechanism of selecting the cells, which have a likelihood of being the most influential faulty cells. This information can be obtained from the testing system as cells are marked with different colors depending on the existence of symptoms of faults. Hence, cell marks are used to guide the search process. During traversing the backward slice, the verification status of cells is used to guide the search to the path where the most influential faulty cell may be located

### 4.1 Search for the most influential faulty cell

During the verification process, cells in a spreadsheet are categorized into three groups. These are cells with symptoms of faults, cells without symptoms of faults, and cells, which are unchecked. Those cells with symptoms of faults are of interest during the process of identifying the most influential faulty cell(s). The search for the most influential faulty cell(s) of a given cell is based on the number of faulty precedents and the contribution of the faulty precedents to incorrect dependents. A cell, which has many faulty precedents, is more likely to contain the most influential cell(s) than the one with few faulty precedents. In addition, a faulty cell, which has more incorrect dependents, is more influential than the one with few incorrect dependents. Furthermore, those faulty cells, which are at a higher level of the data dependency graph (i.e., near the input cells) are more influential than those at the lower level of the data dependency graph. Therefore, correcting faulty cells at the highest level of the data dependency graph may correct cells showing incorrect values (without being actually faulty), which are dependent on the corrected cell thereby reducing the effort of the debugging process.

**4.2 An example**

Let us consider the cost calculation spreadsheet shown in figure 2(a). Figure 2(e) depicts the spreadsheet with those cells, which have symptoms of faults highlighted. Those cells, which have symptoms of faults, are shaded red by the interval-based testing system (dark in this paper) and those without symptoms of faults are shaded yellow (light grey in this paper). For a demonstration purpose, intervals were attached only for cells in column D and hence verification is done only for this column.

Suppose the user wants to trace the most influential faulty cell for the final result of the computation in cell D9. Actually, the first thing to examine is the faulty cell itself. If the formula and the expected interval attached are correct, then the fault is due to referencing a faulty cell. In such cases, we need to trace the source of the fault.

In the example given, the task in column D was to compute the personal cost based on the fixed amount given in cell D4 for each location. However, the user made a referencing error. Cell D4 should have been used as an absolute reference in the formula of D5 as it is copied downward. An error of referencing will have an effect on the copies and not on the source. As a result, the cells D6, D7, and D9 are marked as faulty.

The direct precedents of cell D9 are cells D5, D6, and D7 (see figure 3). The direct precedents fall into the two categories: correct (those without symptoms of faults) and faulty. The faulty category, which contains cells D6 and D7, is the candidate for further investigation. The next task is to identify which one of D6 and D7 contains the most influential faulty cell. Since they have equal verification status, we check the number of their faulty direct precedents and dependents. D6 has no faulty precedents but D7 has one faulty precedent, D6. Therefore, the path to D7 should be followed to locate the most influential faulty cell. Next, we check the faulty precedents of D7. It has only one precedent cell, D6, which is marked as faulty. Among the precedents of D6, there is no faulty precedent.

Therefore, cell D6 is considered to be the most influential faulty cell and as a result, whenever the user requests for the most influential faulty cell for cell D9, the system highlights the cell D6. However, if two or more faulty cells have equal number of faulty direct precedents, then we need to consider also the number of their direct dependents.

While using the number of faulty direct dependents, there are two possibilities to consider.
- the number of faulty direct dependents combined with the number of faulty direct precedents
- the number of faulty direct dependents when the number of faulty direct precedents are equal

| | A | B | C | D | E | F | G |
|---|---|---|---|---|---|---|---|
| 1 | Cost share | | | | | | |
| 2 | | | | | | | |
| 3 | Place | PM | Expense factor | Personal cost | Travel cost | others | Sum |
| 4 | | 48.00 | | 24,000.00 | 2,000.00 | 5,000.00 | 31,000.00 |
| 5 | X | 8.25 | 0.17 | 4,125.00 | 343.75 | | 4,468.75 |
| 6 | Y | 17.25 | 0.36 | 1,482.42 | 718.75 | | 2,201.17 |
| 7 | Z | 22.50 | 0.47 | 694.89 | 937.50 | | 1,632.39 |
| 8 | W | | | | | 5,000.00 | 5,000.00 |
| 9 | | 48.00 | 1.00 | 6,302.31 | 2,000.00 | 5,000.00 | 13,302.31 |
| 10 | | | | | | | |

(a) Cost computation

| | A | B | C | D | E | F | G |
|---|---|---|---|---|---|---|---|
| 1 | Cost share | | | | | | |
| 2 | | | | | | | |
| 3 | Place | PM | Expense factor | Personal cost | Travel cost | others | Sum |
| 4 | | 48 | | 24000 | 2000 | 5000 | =SUM(D4:F4) |
| 5 | X | 8.25 | =B5/B$9 | =D4*C5 | =E$4*$C5 | | =SUM(D5:F5) |
| 6 | Y | 17.25 | =B6/B$9 | =D5*C6 | =E$4*$C6 | | =SUM(D6:F6) |
| 7 | Z | 22.5 | =B7/B$9 | =D6*C7 | =E$4*$C7 | | =SUM(D7:F7) |
| 8 | W | | | | | 5000 | =SUM(D8:F8) |
| 9 | | =SUM(B5:B7) | =SUM(C5:C7) | =SUM(D5:D7) | =SUM(E5:E7) | =SUM(F5:F8) | =SUM(G5:G8) |
| 10 | | | | | | | |

(b) Formula view

| | A | B | C | D | E | F | G |
|---|---|---|---|---|---|---|---|
| 1 | | | | | | | |
| 2 | | | | | | | |
| 3 | | | | | | | |
| 4 | | | | | | | |
| 5 | | | | [4000:4250] | | | |
| 6 | | | | [8500:8780] | | | |
| 7 | | | | [10000:12000] | | | |
| 8 | | | | | | | |
| 9 | | | | [24000:24000] | | | |

(c) Expected spreadsheet

|   | A | B | C | D | E | F | G |
|---|---|---|---|---|---|---|---|
| 1 |   |   |   |   |   |   |   |
| 2 |   |   |   |   |   |   |   |
| 3 |   |   |   |   |   |   |   |
| 4 |   |   |   |   |   |   |   |
| 5 |   |   |   | [4125:4125] |   |   |   |
| 6 |   |   |   | [1437.5:1527.34375] |   |   |   |
| 7 |   |   |   | [3984.375:4115.625] |   |   |   |
| 8 |   |   |   |   |   |   |   |
| 9 |   |   |   | [22500:25030] |   |   |   |

(d) Bounding spreadsheet

|    | A | B | C | D | E | F | G |
|----|---|---|---|---|---|---|---|
| 1  | Cost share |   |   |   |   |   |   |
| 2  |   |   |   |   |   |   |   |
| 3  | Place | PM | Expense factor | Personal cost | Travel cost | others | Sum |
| 4  |   | 48.00 |   | 24,000.00 | 2,000.00 | 5,000.00 | 31,000.00 |
| 5  | X | 8.25 | 0.17 | 4,125.00 | 343.75 |   | 4,468.75 |
| 6  | Y | 17.25 | 0.36 | 1,482.42 | 718.75 |   | 2,201.17 |
| 7  | Z | 22.50 | 0.47 | 694.89 | 937.50 |   | 1,632.39 |
| 8  | W |   |   |   |   | 5,000.00 | 5,000.00 |
| 9  |   | 48.00 |   | 1.00 | 6,302.31 | 2,000.00 | 5,000.00 | 13,302.31 |
| 10 |   |   |   |   |   |   |   |

(e) A spreadsheet with symptom of faults

*Figure 2: Interval-based testing in action*

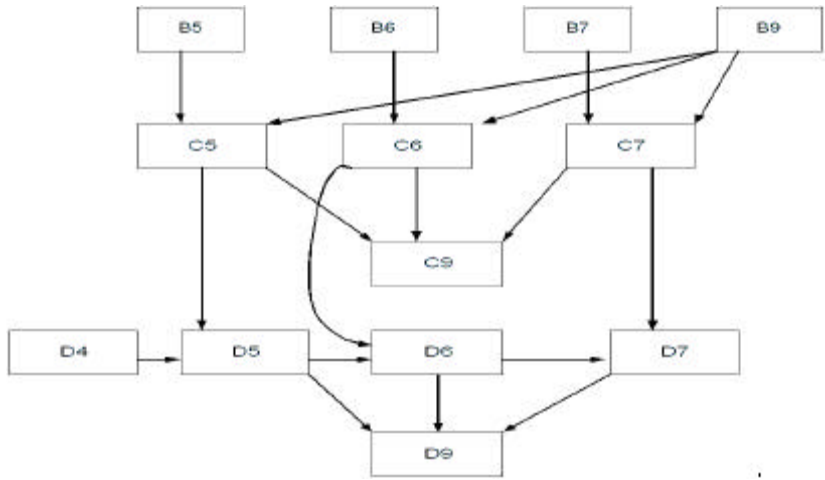

*Figure 3: Data dependency graph for cell D9*

If we use the first choice, we may find the most influential cell without going far in the data dependency graph. This influential cell may not be the most influential for the faulty cell under consideration but it contributes to many other faulty dependent cells. Therefore, correcting this cell may also correct many other cells, which are not in the dynamic backward slice of the cell under consideration. This option identifies the most influential cell in terms of the number of incorrect dependents of a cell. If we use the second choice, then we can reach to the most influential cell with respect to the cell under consideration, which is at a higher level in the data dependency graph. Therefore, correcting this cell may correct many cells in the dynamic backward slice of the cell under consideration. Though both options provide the possibility of correcting many cells, we prefer to locate the most influential cell using the second option as this identifies the most influential cell for the cells in the dynamic backward slice. In the case where two or more faulty cells have equal number of faulty precedents and dependents, one of them will be chosen arbitrarily.

### 4.3 Fault tracing algorithm

The algorithm for identifying the most influential faulty cell is presented in algorithm 1 for propagated faults. Let $C_e$ be the erroneous cell we are interested in to identify it's most influential faulty cell(s).

*Algorithm 1: Algorithm to identify the most influential faulty cell*

1: $GE = \{C_i | C_i$ is a direct precedent of $C_e$ and $C_i$ has a symptom of fault$\}$.
2: If $GE = \emptyset$, then $C_e$ is the most influential faulty cell and stop.
3: Compute the faulty direct precedents of elements of GE.
   $GEE_i = \{C_{ij} | C_{ij}$ is a direct precedent of $C_i$ and $C_{ij}$ has a symptom of fault$\}$.
4: Extract precedents with maximal number of second order precedents
   $GGEE = \{C_i$ such that $|GEE_i|$ is maximal$\}$.
5: If $GGEE$ is singleton, repeat from step 1 with $C_e$ $\hat{I}$ $GGEE$.
6: If $|GGEE| > 1$ compute the faulty direct dependents of $C_i$.
   $GED_i = \{D_{ij} | D_{ij}$ depends on $C_i$ and $D_{ij}$ has a symptom of fault$\}$.
7: Extract precedents with maximal number of second order precedents and maximum number of dependents
   $GED = \{C_i | C_i$ $\hat{I}$ $GGEE$ and $|GED_i|$ is maximal$\}$
8: If $GED$ is singleton, repeat from steps 1 with $C_e$ $\hat{I}$ $GED$.
9: If $|GED| > 1$ take arbitrarily $C_i$ $\hat{I}$ $GED$ and repeat from step 1.

Thus, departing from erroneous cell $C_e$, first a section of the backward slice is computed in such a way that at each step, one aims to reduce a branching slice to the most promising singleton (steps 1 to 5). If this attempt terminates in a situation where the algorithm cannot decide which cell in a set of faulty candidates is most influential, it changes direction and computes the forward slice from the respective dependents, assuming that the root having most dependents is the one the user should look at first.

### 5. CONCLUSION

Debugging involves the identification of the location of the actual faults and fixing the faults given testing revealed some incorrect values. Since most spreadsheets are not produced by software professionals, this process calls for machine support. In spreadsheets, the identification of cells with symptoms of faults and those without symptoms of faults is carried out by the testing system (i.e., interval-based testing).

As spreadsheets are dataflow-driven, faults are propagated in the direction of the dataflow. Therefore, we need a mechanism of identifying the most influential faulty cell in the data dependency graph against the direction of the data flow, so that correcting it may correct many cells in the data dependency graph thereby simplifying the debugging process.

In this paper, we have presented a technique for the identification of the most influential faulty cell(s) for a given faulty cell, which has a propagated fault. Unlike conventional software fault localization techniques, which apply dicing, we do not limit the number of faulty cells in the dynamic backward slice of the cell under consideration to one. Several cells with symptoms of faults can appear in the analog of a dynamic backward slice of a given faulty cell. For the identification of the most influential faulty cells, the fault tracing strategy uses the dataflow information, which is available from the spreadsheet language and the cell marks obtained from the testing system. Path selection is based on the number of faulty precedent and dependent cells.